\documentstyle[twocolumn,seceq,epsbox]{jpsj}

\def\mbf(#1){\mbox{\boldmath $#1$}}
\def\runtitle{Time-reversal symmetry breaking surface states}
\def\runauthor{Kazuhiro {\sc Kuboki} and Manfred {\sc Sigrist}}

\title
{Time-reversal symmetry breaking surface states in $t-J$ model}

\author{Kazuhiro Kuboki$^a$ and Manfred Sigrist$^b$}

\inst{$^a$Department of Physics, Kobe University,
 Kobe 657-8501, Japan \\ 
$^b$ Yukawa Institute for Theoretical Physics, Kyoto University, 
           Kyoto 606-01, Japan}

\abst{  
  Recently a phenomenological Ginzburg-Landau (GL) theory has been proposed to 
  describe the occurrence of a locally time-reversal symmetry (${\cal
T}$) breaking  
  state near a Josephson junction between unconventional 
  superconductors. 
  In this paper we derive this type of GL free energy microscopically
from the $t-J$ model  
  within a slave-boson mean-field approximation. 
 The resulting GL free energy is shown to satisfy the conditions 
  to have a ${\cal T}$-violating surface state. 
  The existence of this junction state may explain some of the recent
  experiments on High-$T_c$ superconductors. }

\kword{unconventional superconductivity, broken time reversal
symmetry}

\begin{document}
\sloppy
\maketitle

  \section{Introduction}

The symmetry of the superconducting state in high-temperature
superconductors (HTSC)
has been a subject of intensive study as an important clue 
to clarify the mechanism of their superconductivity. 
The Josephson effect allows us to investigate directly the phase 
properties of a superconducting order parameter (OP), 
and thus it is a powerful experimental probe for this study. 
Many experiments demonstrate that the superconducting OP 
in these systems has a predominantly $d_{x^2-y^2}$-wave character,
i.e. the OP  changes sign under 90$^\circ$-rotation in the CuO$_2$
plane \cite{TEST,SR}.  

In $d$-wave superconductors interface properties can be qualitatively 
different from those of conventional superconductors 
because of the nontrivial angular dependence of their pair wave functions. 
We have shown that 
a locally time-reversal symmetry (${\cal T}$) breaking state 
can occur near an Josephson junctions between $d$-wave superconductors 
and, in general, unconventional superconductors
\cite{SBL,KS,SKLMR,LT21A}.
This ${\cal T}$-violating state exists only near 
the surface and decays exponentially toward 
the bulk. It has important consequences 
on Josephson effects. 
The arguments which led to this conclusion were based on a 
phenomenological Ginzburg-Landau (GL) theory including several 
assumptions.\cite{SBL,KS,SKLMR,LT21A,note4,note5}

In this paper we derive the GL free energy from the 
$t-J$ model within a slave-boson mean-field approximation, and 
demonstrate that it is possible to have a ${\cal
T}$-violating surface state. 
The reason we consider the $t-J$ model is the following. 
Mean-field (MF) theories of the $t-J$ model based on a slave-boson method 
predict a superconducting 
state with a $d_{x^2-y^2}$-symmetry,\cite{SHF,KL} and they 
may explain the magnetic\cite{TKF,FKNT} as well as 
the transport\cite{NL} properties of HTSC 
if the gauge fields representing the fluctuations around the MF 
solutions are properly taken into account. 
Thus, it is interesting to study whether the $t-J$ 
model leads to a $ {\cal T} $-violating state, in particular,
at the Josephson junction.

\section{Mean Field theory and GL expansion of free energy}

We consider the $t-J$ model on a square lattice 
with the Hamiltonian
\begin{equation} 
  \begin{array}{rl}
  H = & - \displaystyle t \sum_{<i,j> \sigma} 
    ( {\tilde c}_{i,\sigma}^{\dagger} 
        {\tilde c}_{j,\sigma} + h.c.) \nonumber \\ 
     + & \displaystyle J \sum_{<i,j>} {\vec S}_i \cdot {\vec S}_j \,
\end{array}
\end{equation}
where the summation is taken over nearest-neighbor bonds $\langle i,j\rangle$, 
and ${\tilde c}_{i\sigma} \equiv c_{i\sigma}(1-n_{i,-\sigma})$. 
We use the slave-boson method to enforce the condition of no 
double occupancy by introducing spinons ($f_{i\sigma}$; fermion) 
 and holons ($b_i$; boson) (${\tilde c}_{i\sigma} = b_i^\dagger f_{i\sigma}$). 
Then the Hamiltonian is decoupled with the following  
  order parameters (OP)\cite{SHF,KL}:
 (1) the bond OP, 
$\langle b^\dagger_j b_i \rangle \equiv \chi_B$ and 
$\langle f^\dagger_{j\sigma}f_{i\sigma} \rangle \equiv \chi_F$ which
we assume to be homogeneous for all nearest-neighbor bonds;
(2) the OP for the Bose condensation of holons, $\langle b_j b_i
\rangle$ \cite{foot0};
  (3) the singlet RVB OP,
$\langle f^\dagger_{i\uparrow}f^\dagger_{j\downarrow} 
- f^\dagger_{i\downarrow}f^\dagger_{j\uparrow}\rangle \equiv \Delta^{*}_{ij}$. 
The superconducting OP is given by the product of the last two,
$\langle b_j b_i \rangle\Delta^{*}_{ij} $.
  In a slave-boson mean field theory there are four kinds of ordered
states in all of which the bond OP are finite:
  (a) the uniform RVB state where only the bond OPs are finite; 
  b) the spin gap state where the singlet RVB OP is also finite \cite{foot1}; 
  In this state there is a (pseudo-) gap in the spin, but not 
  in the charge excitations. Hence the name spin gap state; 
  c) the superconducting state where all three OP's listed above are finite; 
  d) the Fermi liquid state \cite{foot2}. 
  A schematic phase diagram is shown in Fig.1. 

  
In this paper we consider only the optimally and over doped case where 
  $T_{BE} \geq T_{RVB}$, and so the critical temperature for superconductivity, 
  $T_c$, is given by $T_{RVB}$. 
  (In other words we do not treat the case where the onset of superconductivity 
  is given by the Bose condensation of holons.)
  In this case we can take $\chi_B = \delta$, since 
  we always consider the case $T \leq T_{BE}$.
Since the superexchange interaction exists only 
for nearest-neighbor bonds, the 
$d_{x^2-y^2}$- and the extended $s$-wave are natural candidates 
for the symmetry of the superconducting OP. 
The former (latter) is defined by 
$\Delta_d(i) = (\Delta_{i,i+x}-\Delta_{i,i+y})/2$ 
($\Delta_s(i) = (\Delta_{i,i+x}+\Delta_{i,i+y})/2$). 
Following a standard procedure we expand the free energy with respect to 
$\Delta_d$ and $\Delta_s$.\cite{SigUe} 
  The resulting GL energy is given after taking 
a continuum limit 
\begin{equation} 
  \begin{array}{rl}
F = & \displaystyle \int d^3 x [\sum_{j=d,s} \{ \tilde{a}_j(T) |\Delta_j|^2 
+ \beta_d |\Delta_j|^4 + K_j |{\bf D}
\Delta_j|^2 \} \\ 
& \\ 
&    + \gamma_1 |\Delta_d|^2 |\Delta_s|^2 + \frac{1}{2} \gamma_2
(\Delta_d^{*2} \Delta_s^2 + \Delta_d^2 \Delta_s^{*2}) \\ 
& \\ 
& + \tilde{K} \{ (D_x \Delta_d)^* (D_x \Delta_s) - (D_y \Delta_d)^* (D_y
\Delta_s) \\ & \\ & \displaystyle 
   + c.c. \} + \frac{1}{8 \pi} (\nabla \times {\bf A})^2 ] 
\end{array} 
\end{equation}
where 
${\vec D} = \nabla - i(2\pi/\Phi_0){\vec A}$ with $\Phi_0 (= hc/2e)$ 
being the standard flux quantum. 
The coefficients in $F$ are given as 
\begin{equation}
\begin{array}{rl}
\alpha_j = & \displaystyle \frac{3J}{4}\big( 1 - \frac{3J}{8N} 
  \sum_k \frac{\tanh (\xi_k/2T)}{\xi_k} \omega_j^2(k)\big)
\\
\beta_j = & \displaystyle \Bigl(\frac{3J}{4}\Bigr)^4\frac{1}{N}
  \sum_k I(\xi_k) \omega_j^4(k)
\\
\gamma = & \displaystyle \Bigl(\frac{3J}{4}\Bigr)^4\frac{1}{N}
  \sum_k I(\xi_k) \omega_d^2(k)\omega_s^2(k)
\\
K_j = & \displaystyle W_F^2\frac{9J^2}{32N}
  \sum_k \frac{f"(\xi_k)}{\xi_k} \sin^2 k_x \omega_j^2(k)
\\
{\tilde K} = & \displaystyle W_F^2\frac{9J^2}{32N}
  \sum_k \frac{f"(\xi_k)}{\xi_k} \sin^2 k_x \omega_d(k) \omega_s(k)
\end{array}
\end{equation}
where $j = d$ or $s$, $\gamma_1 = 2\gamma$, $\gamma_2 = \gamma/2$,
  $\omega_d(k) = \cos k_x -\cos k_y$ and $\omega_s(k) = \cos k_x + \cos k_y$. 
Here 
\begin{equation}
\begin{array}{rl}
I(\xi_k) =  & \displaystyle \frac{1}{2\xi^2_k} 
  \Bigl[f'(\xi_k) + \frac{1}{2\xi_k}\tanh(\frac{\xi_k}{2T})\Bigr] 
\\
W_F = & \displaystyle t\delta + \frac{3}{8}J\chi_F
\end{array}
\end{equation} 
  and $f(\xi_k)$ is the Fermi distribution function. 

The surface energy at the junction is calculated under the assumption of 
a specularly reflecting surface. We consider a planar interface parallel 
 to the $c$-axis as shown in Fig.2. 
 In Fig.2 the left and the right hand side are the same $d$-wave 
  superconductors described by the $t-J$ model. 
  Here the crystalline $a$-axis of the left hand side 
  $(L)$ is normal to the interface, while that in the right hand side $(R)$ 
  is taken as a free parameter, 
  denoted as $\varphi$ ($0 \leq \varphi \leq \pi$).  
 (We can treat an $S/D$-junction, where the left side is an $s$-wave 
 superconductor, in a similar way. We consider this case in $\S$ 5.)
  In this configuration the important effects are associated mainly 
  with the OP on the right hand side, so that we will simply represent
 the left hand side by a  
  single $d$-wave order parameter $\Delta_0$ only.  
  
   
  The transmission and the reflection of electrons at the 
  interface ($I$) may be described by the following Hamiltonian, 
  \begin{equation}
  \begin{array}{rl}
  H_I =& \displaystyle \sum_\sigma \sum_{k,p} \big[t_{kp} 
         \big(f_{k\sigma}^{\dagger(L)} f_{p\sigma}^{(R)}
      + f_{p\sigma}^{\dagger(R)} f_{k\sigma}^{(L)} \big) \nonumber \\
  & \\
      +& \displaystyle r_{kp} \big(f_{k\sigma}^{\dagger(R)} f_{p\sigma}^{(R)} 
                  + f_{p\sigma}^{\dagger(R)} f_{k\sigma}^{(R)}\big)\big] \
  \end{array}                 
  \end{equation}
  where $f^{(L)}_{k\sigma}$ ($f^{(R)}_{k\sigma}$) is the spinon operator for 
  the left (right) side, and the matrix elements for tunneling ($t_{kp}$) 
  and the reflection ($r_{kp}$) are taken to be real.   
  Treating $H_I$ in a second-order perturbation theory we get 
  the surface free energy $F_I$ to lowest order in $\Delta$'s,

\begin{equation} 
\begin{array}{rl}
F_{I} =& \displaystyle \int_{I} dS \big[ \sum_{i,j=\{d,s\}} g_{ij} 
      (\varphi) \Delta^*_i \Delta_j \nonumber \\
& \\
      +& \displaystyle \sum_{i=\{d,s\}} t_i (\varphi) 
        (\Delta^*_0 \Delta_i + \Delta_0 \Delta^*_i)\big]. \
\end{array}
\end{equation}
The first term originates from the reflection of the Cooper pairs at the 
interface and the second term represents the coupling between 
the two sides ($g_{ij} = g_{ji}$).
  For the $D/D$-junction composed of the same superconductors, 
  the coefficients in eq.(2.6) are given as
  \begin{equation}
  \begin{array}{rl}
  t_d = & \displaystyle \Bigl(\frac{3J}{4}\Bigr)^2 
          \sum_{kp} t_{kp}^2 J_1(\xi_k,\xi_p) \omega_d(k)\omega_d(p)    
  \\
  t_s = & \displaystyle \Bigl(\frac{3J}{4}\Bigr)^2 
          \sum_{kp} t_{kp}^2 J_1(\xi_k,\xi_p) \omega_d(k)\omega_s(p)          
  \\    
  g_d = & \displaystyle \Bigl(\frac{3J}{4}\Bigr)^2 
       \sum_{kp} \big[r_{kp}^2 J_1(\xi_k,\xi_p) \omega_d(k)\omega_d(p) \\
  & \\    
    + & \displaystyle (r_{kp}^2 + t_{kp}^2) J_2(\xi_k,\xi_p) \omega_d(k)^2 \big]
  \\ & \\
  g_s = & \displaystyle \Bigl(\frac{3J}{4}\Bigr)^2 
       \sum_{kp} \big[r_{kp}^2 J_1(\xi_k,\xi_p) \omega_s(k)\omega_s(p) \\
  & \\
    + & \displaystyle (r_{kp}^2 + t_{kp}^2) J_2(\xi_k,\xi_p) \omega_s(k)^2 \big]
  \\ & \\
  g_{ds} = & \displaystyle \Bigl(\frac{3J}{4}\Bigr)^2 
       \sum_{kp} \big[r_{kp}^2 J_1(\xi_k,\xi_p) \omega_s(k)\omega_d(p) \\
  & \\
    + & \displaystyle (r_{kp}^2 + t_{kp}^2) J_2(\xi_k,\xi_p) 
    \omega_d(k)\omega_s(k) \big] 
  \end{array}
  \end{equation}  
  with
  \begin{equation}
  \begin{array}{rl}
  J_1(\xi_k,\xi_p) = & \displaystyle \frac{1}{\xi_k^2-\xi_p^2} 
          \big(\frac{\tanh(\frac{\xi_k}{2T})}{2\xi_k}
              -\frac{\tanh(\frac{\xi_p}{2T})}{2\xi_p}\big)
  \\ & \\
  J_2(\xi_k,\xi_p) = & \displaystyle \frac{2\xi_k}{\xi_k-\xi_p}I(\xi_k) \\
  & \\
      + & \displaystyle \frac{\xi_p}{(\xi_k+\xi_p)(\xi_k-\xi_p)^2}
        \big(\frac{\tanh(\frac{\xi_k}{2T})}{\xi_k} 
           - \frac{\tanh(\frac{\xi_p}{2T})}{\xi_p}\big).
  \end{array}
  \end{equation}   
  The $J_1$ terms represent the usual tunneling and the reflection processes of 
  a Cooper pair. On the other hand, the $J_2$ terms give different types
  of reflection processes where one of the particle consisting of a 
  Cooper is reflected at the interface, while the other one tunnels to 
  the opposite side (see in the Appendix for the derivation of the
surface terms and the interpretation of $ J_2 $). 
 
    Here we follow the method of Ref. \cite{Bruder,Barash} 
  in taking the angular dependences of $t_{kp}$ and 
  $r_{kp}$, which are consistent with the assumption of the specularly 
  reflecting surface, 
  \begin{equation}
\begin{array}{rl}
t_{kp}^2 = & \displaystyle {\tilde t}^2 \delta(k_\parallel-p_\parallel)
             \delta(k_\perp-p_\perp)\frac{v_\perp(p)}{v(p)}\theta(v_\perp(p))
\\
r_{kp}^2 = & \displaystyle {\tilde r}^2 \delta(k_\parallel-p_\parallel)
             \delta(k_\perp+p_\perp)\frac{v_\perp(p)}{v(p)}\theta(v_\perp(p))
\\
\end{array}
\end{equation}  
  where $\theta$ is the step function. 
  It should be noted here that if the angular dependences of the tunneling 
  matrix elements were omitted,
  the coupling between two superconductors should vanish, 
and thus it would lead to unphysical results.

  \section{Relation to Phenomenological Theory}
  
  Now let us briefly summarize the content of the phenomenological theory 
  of Ref.4 to describe a ${\cal T}$-violating surface state.
  The properties of the bulk system are described by the free energy $F$.
  We assume that only a $d$-wave OP is 
  present in the bulk system, and the coupling between $\Delta_d$ and 
  $\Delta_s$ is repulsive, if the latter were present. 
  Namely, $a_d$ is negative below a critical 
  temperature, while $a_s > 0$ for all $T$, and $\gamma_1 - \gamma_2 > 0$ and 
  $\gamma_2 > 0$. 
  An important point here is that positive $\gamma_2$ favors the 
  relative phase 
  between $\Delta_d$ and $\Delta_s$, denoted as $\phi_{ds}$, to be  
  $\pm\pi/2$, i.e., the system would break ${\cal T}$ if both $d$ and $s$-
  components coexist.
  We expect $\gamma_2 > 0$ for the following reason. 
  The complex combination $d \pm is$, resulting from $ \gamma_2 >0 $,
  would open a complete gap in the excitation spectrum, 
  such that the system would gain more condensation energy. 
  If $\phi_{ds} =0$ or $\pi$ (($d \pm s$)-state), 
  the nodes remain though their locations are shifted,  
  and the gain of the condensation energy would be smaller.
  
  The interface properties are described by $F_I$.
  In the presence of an interface, $\Delta_s$ can be induced by the reflection
  of Cooper pairs and by the proximity effect from $\Delta_0$ in $(L)$.  
  These processes are possible because of the lowering of the symmetry 
  (translational and point-group) in the presence of the interface.
  In usual cases $\phi_{ds}$ is determined by the bilinear coupling terms in 
  $F_I$ which induces the finite $\Delta_s$. On the other hand, the
  $\gamma_2$ term favoring  
  $\phi_{ds} = \pm \pi/2$ is biquadratic and it would not be dominant 
  in the sense of GL theory.
  Then $\phi_{ds}$ takes a value 0 or $\pi$ depending on the sign
  of $t_d$, $t_s$ and $g_{ds}$. There is no ${\cal T}$-breaking in this case. 
  Under certain conditions, however, a different relative phase is favored 
  leading to a state which breaks ${\cal T}$. 
  In Ref.4 it was argued based on the symmetry consideration that $t_d$ and 
  $g_{ds}$ should vanish at $\theta = \pi/4$ and are small for $\theta$ 
  close to $\pi/4$, while $t_s$ remains always finite. 
  Then, for $\theta \sim \pi/4$ 
  $\phi_{ds}$ is determined by $\gamma_2$, and, thus,
  it is possible to have a locally 
  ${\cal T}$-violating state near the interface.   

  We numerically calculate coefficients in $F$ and $F_I$ as functions of 
 the doping rate $\delta$ and the temperature
  $T$. We take $t/J= 3$ throughout in this 
  paper. First we solve the self-consistency equations for the uniform RVB state
  (the state where only $\chi_F$ and $\chi_B = \delta$ are finite), 
  to get $\chi_F$ and the chemical potential, $\lambda_F$.
  By using them we calculate the coefficients in $F$ and $F_I$. 
  For the doping rate where the superconductivity is observed in 
  real systems,  i.e., $0 < \delta < 0.3$, only $\alpha_d$ can be negative 
  and all other coefficients of $F$ are positive definite \cite{foot3}. 
  This implies that we deal with a single $d$-wave component. 
  We find indeed a positive
  $\gamma_2$ 
  which favors the phase difference $\phi_{ds} = \pm \pi/2$. 
  Hence, the above results are consistent with the assumptions of the 
  previous purely phenomenological theory. 
  
  Next we consider the interface terms $F_I$. By taking the angular
  dependences of $t_{kp}$ and 
  $r_{kp}$ as in eq.(2.9), we get the the coefficients in $F_I$ as functions of 
  $\varphi$, i.e., the angle between the interface and the crystal $a$-axis 
  of the right hand side. 
  We can explicitly examine the following properties using the expressions 
  of $t_i$ and $g_{ij}$ ($i, j = d, s$):
  \begin{equation}
  \begin{array}{rl}
  t_s(\varphi \pm \pi/2) = & t_s(\varphi), \ \ 
  t_d(\varphi \pm \pi/2) =  -t_d(\varphi)
  \\
  t_s(\varphi \pm \pi) = & t_s(\varphi), \ \ 
  t_d(\varphi \pm \pi)  =  t_d(\varphi)
  \\
  t_s(-\varphi) = & t_s(\varphi), \ \ 
  t_d(-\varphi)  =  t_d(\varphi)
  \end{array}
  \end{equation}
  and 
  \begin{equation}
  \begin{array}{rl}
  g_{ii}(\varphi \pm \pi/2) = & g_{ii}(\varphi), \ \ 
  g_{ds}(\varphi \pm \pi/2) = -g_{ds}(\varphi) 
  \\
  g_{ii}(\varphi \pm \pi) = & g_{ii}(\varphi), \ \ 
  g_{ds}(\varphi \pm \pi) =  g_{ds}(\varphi)
  \\
  g_{ii}(-\varphi) = & g_{ii}(\varphi), \ \ 
  g_{ds}(-\varphi) =  g_{ds}(\varphi)  
  \end{array}
  \end{equation}
  where $i = d, s$.  
  We have calculated $t_i$ and $g_{ij}$ numerically, as shown in
Fig. 3, with a choice of parameters $\delta = 0.15$ and $T = 0.8T_c$ 
  ($T_c \sim 0.108J$).
  The important point here is that $t_s$ does not vanish for any 
  value of $\varphi$. This property is robust for any doping
  $\delta$ and temperature $T$. 
  Another point is that 
  $|t_s|$ is much smaller than $|t_d|$ except very near $\varphi = \pi/4$.
  This can be explained as follows.
  The integrand in the expression of $t_s$ has a factor 
  $(\cos k_x + \cos k_y)$, while that of $t_d$ has $(\cos k_x -\cos k_y)^2$.
  Since we are treating the square lattice system, the factor 
  $(\cos k_x + \cos k_y)$ is small everywhere on the Fermi surface, 
  for the doping rate as small as $\delta = 0.15$,
  while the factor $(\cos k_x -\cos k_y)^2$ can be large there. 
  Hence $|t_d|$ can be much larger than $|t_s|$. (For $\varphi = \pi/4$,  
  $t_d$ has to vanish by symmetry. Then $|t_s|$ can be larger than $|t_d|$ 
  if $\varphi$ is close enough to $\pi/4$.)



  \section{${\cal T}$-breaking state and Surface Current} 
   
  In this section we analyze the appearance of the ${\cal T}$-breaking
state.
  For this purpose we minimize 
  the total free energy, $F_{tot} = F + F_I$, with respect to the OP,
$\Delta_s $ and $ \Delta_d $, and the vector potential ${\vec A}$. 
  We use a coordinate system $({\tilde x}, {\tilde y})$, with 
  ${\tilde x}$ (${\tilde y}$) being perpendicular (parallel) to the 
  interface. 
  Under the transformation from $(x, y)$ to $({\tilde x}, {\tilde y})$, 
  only ${\tilde K}$ terms are changed, and it transforms to 
  \begin{equation}
  \begin{array}{rl}
  & \displaystyle {\tilde K} \big[ \cos 2\varphi \big\{ 
    (D_{\tilde x}\Delta_d)(D_{\tilde x}\Delta_s)^{*}
    -(D_{\tilde y}\Delta_d)(D_{\tilde y}\Delta_s)^{*} \big\} \\
  & \\
  & \displaystyle + \sin 2\varphi \big\{
    (D_{\tilde x}\Delta_d)(D_{\tilde y}\Delta_s)^{*}
    +(D_{\tilde y}\Delta_d)(D_{\tilde x}\Delta_s)^{*} \big\} \big].
  \end{array}
  \end{equation}
  Since there is no spatial variation of the OP along 
  ${\tilde y}$-direction, 
  the GL equations are formally written as
  \begin{equation}
  \displaystyle \frac{\partial F}{\partial \Delta_i({\tilde x})} 
  = \partial_{\tilde x} \frac{\partial F}{\partial (\partial_{\tilde x}
  \Delta_i({\tilde x}))} \ \ (i = d, s)
  \end{equation}
  and 
  \begin{equation}
  \begin{array}{rl}
  J_{\tilde x} \equiv & \displaystyle 
  -\frac{\partial F}{\partial A_{\tilde x}} = 0, \\
  & \\
  J_{\tilde y} \equiv & \displaystyle 
  -\frac{\partial F}{\partial A_{\tilde y}} =
  -\frac{1}{4\pi}\partial_{\tilde x}B
  \end{array}
  \end{equation}
  where $B = \nabla_{\tilde x}A_{\tilde y} - 
  \nabla_{\tilde y}A_{\tilde x}$ 
  and $A_{\tilde x} = A_x\cos\varphi - A_y\sin\varphi , 
  A_{\tilde y} = A_x\sin\varphi + A_y\cos\varphi$. 
  The boundary conditions at the interface are  
  \begin{equation}
  \displaystyle \big(\frac{\partial F}{\partial (\nabla_{\tilde x} 
  \Delta_i({\tilde x}))}
  + \frac{\partial F_I}{\partial \Delta_i({\tilde x})}\big) 
  \biggl\vert_{{\tilde x}=0} = 0 \ \ (i = d, s)
  \end{equation}
  and 
  \begin{equation}
  B({\tilde x} = 0) = 0. 
  \end{equation} 
  
Now we analyze the instability to a ${\cal T}$-violating state 
  at the temperature $T^{*}$. We consider a junctions as described in
Fig.2, and assume that on both sides the superconductors have the same
properties, in particular, the same $ T_{cd} $.
  At $T = T_c$, $\Delta_d$ and $\Delta_0$ ($d$-wave) get finite, 
  and $\Delta_s$ is induced simultaneously near the interface due to 
  the $t_s$ and $g_{ds} $ term. 
  For $\varphi = \pi/4$, $t_d$ and $g_{ds}$ vanish, so only the
$\gamma_2$term determines the relative phase $\phi_{ds}$. 
  Since $\gamma_2$ ($>0$) favors $\phi_{ds} = \pm \pi/2$, ${\cal
T}$-breaking occurs. Thus for 
  $\varphi = \pi/4$, we have $T^{*} = T_c$, namely,  
  ${\cal T}$-violation should occur at $T_c$, the bulk 
  superconducting transition temperature, irrespective of any other
details. 
  
  Numerical calculation for the GL equations confirms this argument. 
  We have solved GL equations (4.2) and (4.3) under the boundary
conditions, 
  eq. (4.4) and (4.5).  
  In Fig.4 we show the spatial dependences of $\Delta_d$, $\Delta_s$ 
  and $\phi_{ds}$ for $\delta = 0.15$, i.e., so-called optimum doping. 
  The results for other values of $\delta$ is qualitatively the same.
  For $\varphi = \pi/4$, $\phi_{ds}$ always takes a 
  value $\pi/2$ or $-\pi/2$, once $T$ becomes lower than $T_c$.  
  As $\varphi$ moves away from $\pi/4$, $t_d$ and $g_{ds}$ become finite
  and compete with $\gamma_2$, which favors ${\cal T}$-breaking states. 
    Since $t_d$ grows very rapidly as $\varphi$ moves away from $\pi/4$,
  $T^{*}$ drastically decreases as a function of $|\varphi - \pi/4|$
  (see Fig.5.).  
    
  However, we should note here that at very low temperature the region
of  ${\cal T}$-violation can be much larger 
  because of the following reason.
  If $T$ is lower than $T_{cs}$, there is a chance 
  to have finite $\Delta_s$ without using tunneling terms. In this case 
  both $t_d$ and $t_s$ can be smaller than 
  $\gamma_2$, with keeping $\Delta_s$ finite \cite{Ogata}.
  Unfortunately $T_{cs}$ derived from the $t-J$ model is quite low
  (of the order of $10^{-3}J$ corresponding to $ T_{cs} / T_{cd} \sim
10^{-2}$), so the region $T < T_{cs}$ is not accessable by the GL theory.

  It has been shown that in a ${\cal T}$-breaking state the surface current 
  along the interface can flow.\cite{note4,Vol,SRU} 
  The condition for the minimum free energy 
  requires the vanishing of the current normal to the junction 
  (i.e., the first equation of (4.3)). This condition in turn leads to a finite 
  current along the junction which induces a magnetic field.
  In Fig.6 the spatial distributions of the surface current and the magnetic 
  field are shown. The extension of the current and the magnetic field are 
  of the order of the penetration depth $\lambda$. Note that no net
current is present, i.e. the integrated current vanishes
\cite{Ohashi},
  \begin{equation}
  \int_0^L d{\tilde x} J_{\tilde y}({\tilde x}) 
  = \frac{1}{4\pi} \big[B({\tilde x} = 0) - B({\tilde x} = L)\big] = 0
  \end{equation} 
  where $L$ is the length of the system ($L \gg \lambda$), and we have 
  used eq.(4.5)
  . 
 
   \section{Case of $S/D$-junctions} 
   
  So far we have considered only $D/D$-junctions, where both sides are 
  the same $d$-wave superconductors.  
  The $S/D$-junction (where the left side is replaced by a different 
  superconductor with isotropic $s$-wave symmetry)
  can be treated similarly.
 
  We consider a conventional superconductor for the left side of the junction. 
  We take the band width, the chemical potential and the magnitude of the order 
  parameter in (L) to be the same as in (R) for simplicity. 
  Our aim here is to compare the qualitative features of $D/D$- and 
  $S/D$-junctions, but not the quantitative comparison.
    
  The surface GL energy can be calculated in a similar way  as in $\S$ 2.
  The modified expressions are simply given by the replacement
  \begin{equation}
  \omega_d(k) \to 1
  \end{equation} 
  in eq.(2.7) for $t_d$ and $t_s$.   
  For the $S/D$-junction $|t_s|$ is larger than $|t_d|$ (for $\varphi$ 
  not so far apart from $\pi/4$), in contrast to the case 
  of $D/D$-junctions. (Fig. 7) 
  However, the value of $|t_s|$ in this case is much 
  smaller than that of $|t_d|$ for $D/D$-junction. 
  This is due to the same reason as we have $|t_d| \gg |t_s|$ for 
  the $D/D$-junction: the factor ($\cos k_x + \cos k_y$) is always small 
  on the Fermi surface, and so $|t_s|$ is reduced even if the OPs 
  of both sides have $s$-wave symmetry. 

  The phase diagram of surface states for $S/D$-junction is shown as a 
  solid line in Fig.5. In Fig.8 the $\varphi$-dependence of the 
  relative phases between $\Delta_d$ and $\Delta_s$ ($\Delta_0$), 
  $\phi_{ds}$ ($\phi_0$) is also shown. 
  We find that in the phase diagram the region of
  ${\cal T}$-breaking states is larger than for the $D/D$-junction. 
  This is the case even when we assume a rather small value of 
  $|\Delta_0|/|\Delta_d^{bulk}|$. 
  In $S/D$-junction $|\Delta_s|$ can be sizable due to large $|t_s|$, 
  and then the $\gamma_2$-term (fourth order) which favors ${\cal T}$-violating 
  states can compete with the second order terms ($t_d$ and $g_{ds}$) even 
  when $\varphi$ is not so close to $\pi/4$. 
  In the case of $D/D$-junction, however, $t_d$ 
  and $g_{ds}$ become dominant once $\varphi$ moves away from $\pi/4$, 
  leading to a very limited region of ${\cal T}$-violating states.

  \section{Conclusions}
  
   We have discussed the S/D- and D/D-interface states using a
GL-formulation derived from the slave-boson mean-field approximation
for the $t-J$ model. The ${\cal T}$-violating interface state was found 
in both cases for certain orientations in agreement with previous
phenomenological studies. 

In addition we found that the ${\cal T} $-violating state is more
likely to occur on an S/D- than on a D/D-interface. This distinction
is mainly due to the fact that the Josephson coupling to a $d$-wave
superconductor is generally weaker, because of the angular phase
structure of the pair wave function which leads to destructive
interference in the tunneling. Furthermore, the shape of the Fermi
surface plays an important role, in particular, for the extended
$S$-wave state. 

Within the GL theory the range of angles $ \varphi $ where the $ {\cal 
T} $-violating state can appear is rather small for the D/D-junction. 
However, it should be noted that for temperatures below $T_{cs}$ this
region in the phase diagram could extend to a wider range. Obviously,
the GL formulation does not allow us to access this temperature
region. In our approach the ratio $ T_{cs} / T_{cd} $ is of the order
$ 10^{-2} $ so that temperatures below $ T_{cs} $ are definitely far
away from the range of validity of a GL theory. The low-temperature
region can only be treated by methods based on Bogolyubov-de Gennes
equations or quasiclassical theory which are considerably more
complicated if one intends to include more realistic microscopic details.

Previously a series of phenomena were discussed,
in connection with broken time reversal symmetry on
interfaces. A first example are flux lines on the
interface which do enclose neither integer nor half-integer multiples
of the standard flux quanta, but some intermediate fractional fluxes
\cite{SBL,KS,YB}. 
Although the observation of fractional flux by Kirtley et
al. \cite{Kirt} is entirely compatible with our discussion here, it is
not clear whether the data could not be explained in an alternative
way. The main problem for this kind 
of experiments lies in the requirement of comparatively long
homogeneous interfaces, a condition hard to satisfy with present
technology. On the other hand, it was also discussed that phase
slip effects on short interfaces could be used as a test
\cite{Kuklov}. So far no experimental data are available for this type 
of effect. Finally an important aspect of the $ {\cal T} $-violating 
state is the presence of spontaneous currents. Their magnitude is
small, however, and together with screening effects they would not
lead to a net magnetization. Thus only a very sensitive probe with
high spatial resolution, smaller or of order London penetration depth, 
would be sufficient to observe this effect. Until now the only 
method which has successfully observed the small magnetic fields induced by
a $ {\cal T} $-violating state are muon spin rotation measurements in zero
external
field \cite{Amato}. 

It is important to notice, however, that also surfaces of d-wave
superconductors can yield states with locally broken time reversal
symmetry. Recent experiments indicate that this type of state
might be realized at low temperatures on YBCO surfaces with [110]-orientation
\cite{Greene}. 
The evidence is given by the splitting of the zero-bias anomaly in
the $IV$-characteristics as discussed by Fogelstr\"om et
al. \cite{Sauls}. Clearly 
similar effects due to the rearrangement of quasiparticle states are
also expected in $ {\cal T} $-violating interfaces as discussed by
Huck et al. and could possibly be tested by spectroscopy with a
scanning tunneling microscope \cite{Huck}.

  \section*{Acknowledgment}

  We are grateful to P.A Lee, T.M. Rice, H. Fukuyama and M. Ogata for
  stimulating discussions.  
  K.K was supported by Grant-in Aid for Scientific Research on Priority 
  Areas "Anomalous metallic state near the Mott transition" from 
  the Ministry of Education, Science and Culture of Japan.

 \appendix
 \section{Derivation of surface terms}
 The surface free energy $F_I$ in $\S2$ is derived in the following way. 
 We describe the transmission and and the reflection of electrons at the 
 interface by the Hamiltonian $H_I$ in eq.(5). 
 Then the total Hamiltonian is given by $H + H_I$. 
 Second order perturbation theory gives the excess energy due to $H_I$:
 \begin{equation}
  \Delta F= \displaystyle -\frac{1}{2} \int_0^\beta d\tau 
  \big\langle T_\tau H_I(\tau)H_I(0) \big\rangle_0 
 \end{equation}
 where $\langle \cdots \rangle_0$ denotes the average with respect to $H$.
 We substitute eq.(5) into this expression and decouple it in terms of 
 Green's functions of spinons for both sides 
 \begin{equation}
 \begin{array}{rl} 
 \Delta F = & \displaystyle \Delta F_1 + \Delta F_2 \\ 
 & \\
 \Delta F_1 = & \displaystyle -\sum_{kp} T\sum_{\epsilon_n} 
   \big[ t_{kp}^2  \big\{ G^L_{21}(p,i\epsilon_n) G^R_{12}(k,i\epsilon_n) \\
 & \\
   + & \displaystyle G^L_{12}(p,i\epsilon_n) G^R_{21}(k,i\epsilon_n) \big\}
   + r_{kp}^2 G^R_{21}(p,i\epsilon_n) G^R_{12}(k,i\epsilon_n) \big] \\
 & \\
 \Delta F_2 = & \displaystyle -\sum_{kp} T\sum_{\epsilon_n} 
   \big[ t_{kp}^2 \big\{ G^L_{11}(p,i\epsilon_n) G^R_{11}(k,i\epsilon_n) \\
 & \\
    + & \displaystyle G^L_{22}(p,i\epsilon_n) G^R_{22}(k,i\epsilon_n) \big\} \\
& \\
  + & \displaystyle \frac{1}{2} r_{kp}^2 \big\{ 
      G^R_{11}(p,i\epsilon_n) G^R_{11}(k,i\epsilon_n)
    + G^R_{22}(p,i\epsilon_n) G^R_{22}(k,i\epsilon_n)  \big\} \big]         
 \\
 \end{array}
 \end{equation}
 where the Green's functions are defined by 
 \begin{equation}
 \begin{array}{rl}
 G^A_{11}(k,i\epsilon_n) = & \displaystyle -\frac{i\epsilon_n+\xi_k}
   {\epsilon_n^2+\xi_k^2+|\Delta_k^A|^2} \\
 & \\
 G^A_{22}(k,i\epsilon_n) = & \displaystyle -\frac{i\epsilon_n-\xi_k}
   {\epsilon_n^2+\xi_k^2+|\Delta_k^A|^2} \\
 & \\
 G^A_{12}(k,i\epsilon_n) = & \displaystyle -\frac{\Delta_k^A}
   {\epsilon_n^2+\xi_k^2+|\Delta_k^A|^2} \\
 & \\
 G^A_{21}(k,i\epsilon_n) = & \displaystyle G^A_{12}(k,i\epsilon_n) ^* 
 \\
 \end{array}
 \end{equation}
 with $A = R$ or $L$, $\epsilon_n = \pi T(2n+1)$, 
 $\Delta_k^R = (3J/4) (\Delta_d\omega_d(k) + \Delta_s\omega_s(k))$ and 
 $\Delta_k^L = (3J/4) \Delta_0\omega_d(k)$. 
 Now we extract the lowest order ($O(\Delta^2)$) terms. In $\Delta F_1$ 
 there are terms of the form $\Delta^L \Delta^R$ and $\Delta^R \Delta^R$  
 in the numerator. The former results in the $t_i$-terms, and the latter 
 leads to the part of the $g_{ij}$-terms. 
 We can also obtain the $O(\Delta^2)$ terms from the denominators of  
 $G_{11}$ and $G_{22}$ in $\Delta F_2$ . 
 These terms are usually discarded in the discussion of the Josephson 
 effect, since it does not depend on the phase difference of 
 $\Delta$'s if both sides of the superconductors have only one component of 
 the order parameter. 
 In the present case, however, this term depends on the phase difference 
 of $\Delta_d$ and $\Delta_s$, and thus it is equally important as the 
 one from $\Delta F_1$. 
 Neglecting terms independent of $\Delta$, we perform the 
 $\epsilon_n$-summation in eq.(A2) to get the following expressions 
 \begin{equation}
 \begin{array}{rl}
 \Delta F_1 = & \displaystyle -\sum_{kp} J_1(\xi_k,\xi_p) \big[
   t_{kp}^2 \big\{(\Delta^L_p)^* \Delta^R_k + \Delta^L_p (\Delta^R_k)^*\big\} 
 \\ & \\
  + & \displaystyle r_{kp}^2 (\Delta^R_p)^* \Delta^R_k \big] \\
 & \\
 \Delta F_2 = & \displaystyle \sum_{kp} J_2(\xi_k,\xi_p) 
  (t_{kp}^2 + r_{kp}^2) |\Delta_p^R|^2 
 \\ 
 \end{array}
 \end{equation}
 The expressions for $t_i$ and $g_{ij}$ (eq.(7) in $\S2$) 
 can be obtained from eq.(A4).  
 For example, $g_d$ is given by the coefficients of $|\Delta_d|^2$ 
 in $\Delta F = \Delta F_1 + \Delta F_2$. 
 
 The meaning of the $J_2$-terms is now obvious. Since we got these terms 
 by picking up $|\Delta^R_p|^2$ in $G_{11}^R$ and $G_{22}^R$ (diagonal 
 components of the Green's functions), only one particle is transfered from 
 (L) to (R) (or (R) to (L)) in this process. This means that one of the 
 electrons consisting of a Cooper pair tunnels to the other side, while 
 the other one is reflected when the Cooper is scattered at the interface. 
 Hence the process including $t_{kp}$ (tunneling matrix element) can lead to 
 the suppression ($g_d$ and $g_s$) and the interference ($g_{ds}$) of the 
 superconducting order parameters.




\bigskip
\bigskip

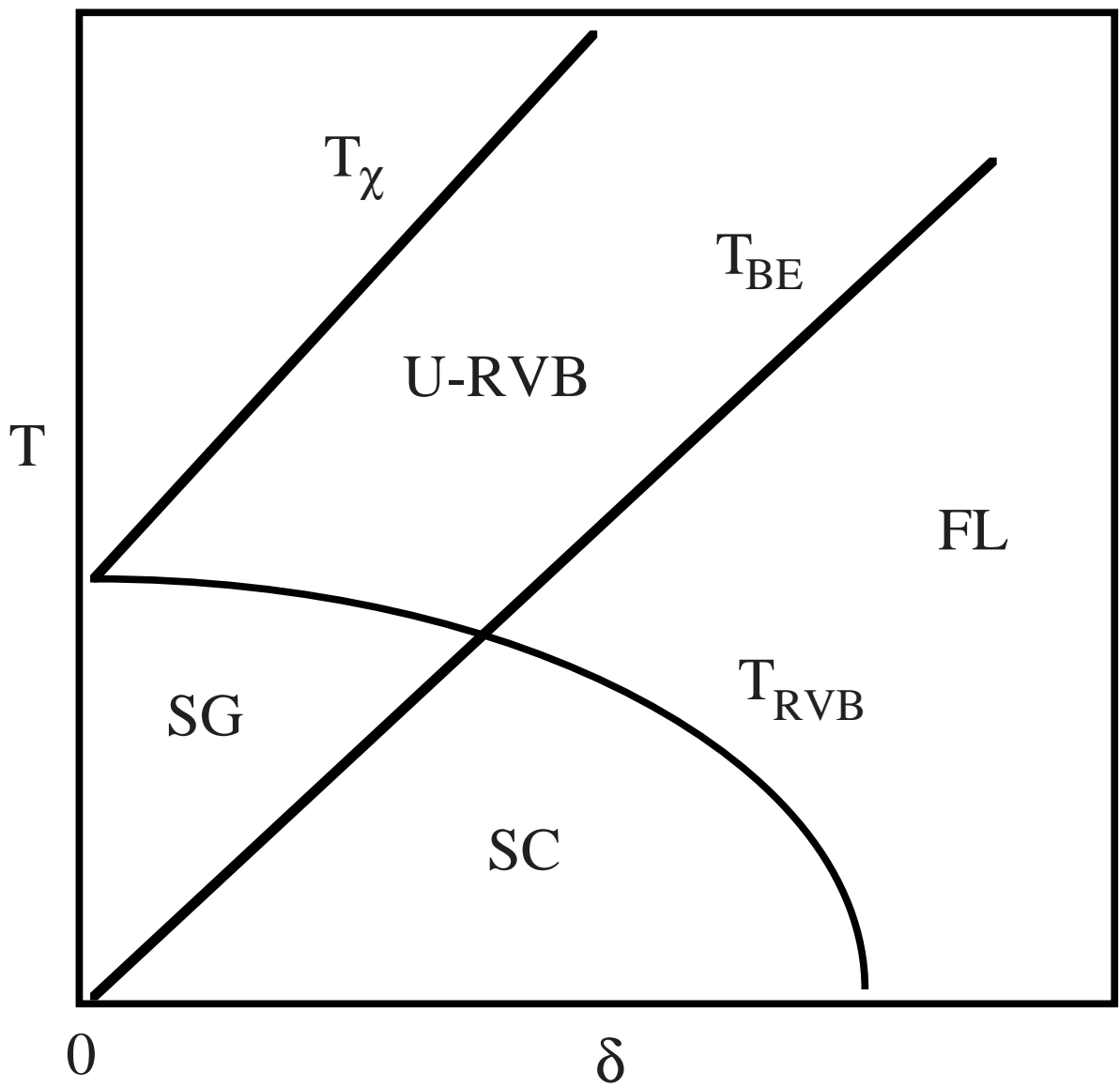
  {\bf Fig. 1.}  
  Schematic phase diagram of the $t-J$ model within a slave-boson 
  mean-field approximation. 
  $T_\chi$, $T_{RVB}$ and $T_{BE}$ are the transition temperature for the 
  bond order, singelt RVB order and the Bose condensation, respectively.

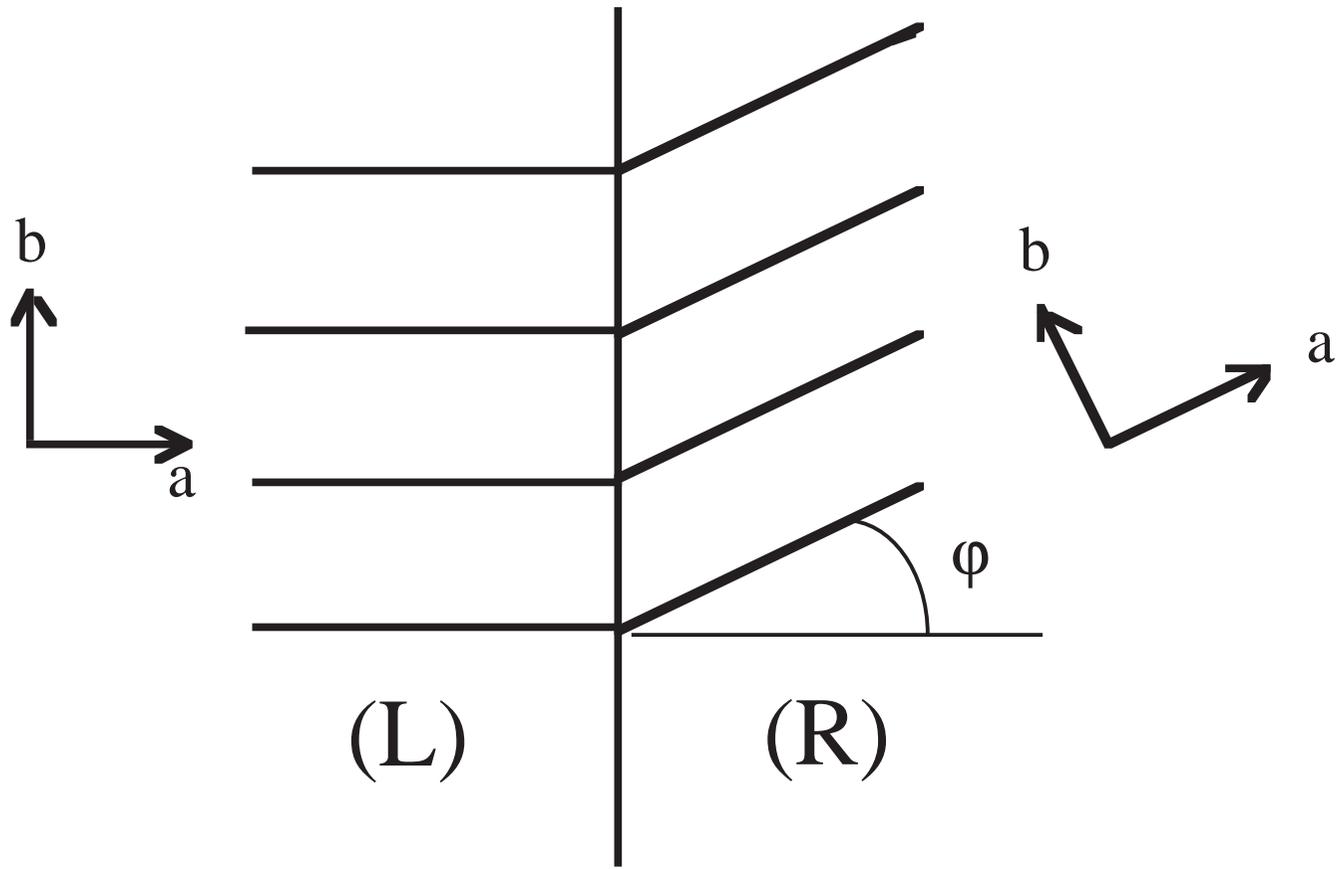
  {\bf Fig. 2.}   
  Interface between $d-$wave superconductors. The lines on both 
  sides indicate the crystalline $a$-axis (the $c$-axis points out of the 
  plane).

  {\bf Fig. 3.}  
  Numerical results of the coefficients in $F_I$ for $D/D$-junction:
  (a) $t_d$ and (b) $g_{ij}$ ($i,j = d,s$).
  Here $\delta = 0.15$, $T = 0.8T_c$, ${\tilde t}^2 = 8.0$ 
  and ${\tilde r}^2 = 0.2$.
  
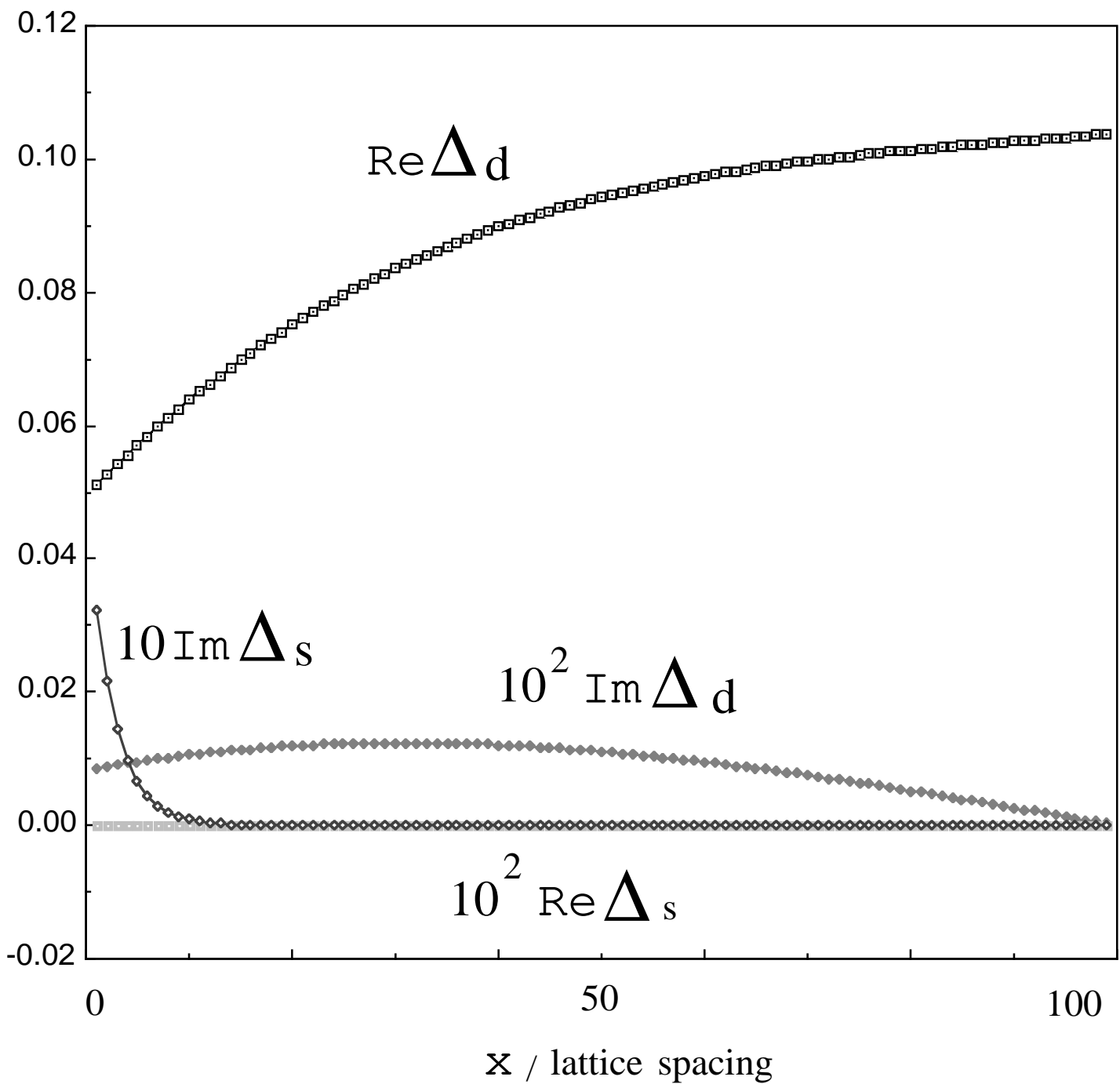
  {\bf Fig. 4}
  Spatial variation of the order parmeters and their relative phase. 
  Here $\delta = 0.15$, ${\tilde t}^2 = 8.0$ and ${\tilde r}^2 = 0.2$
  and $\varphi = \pi/4$.

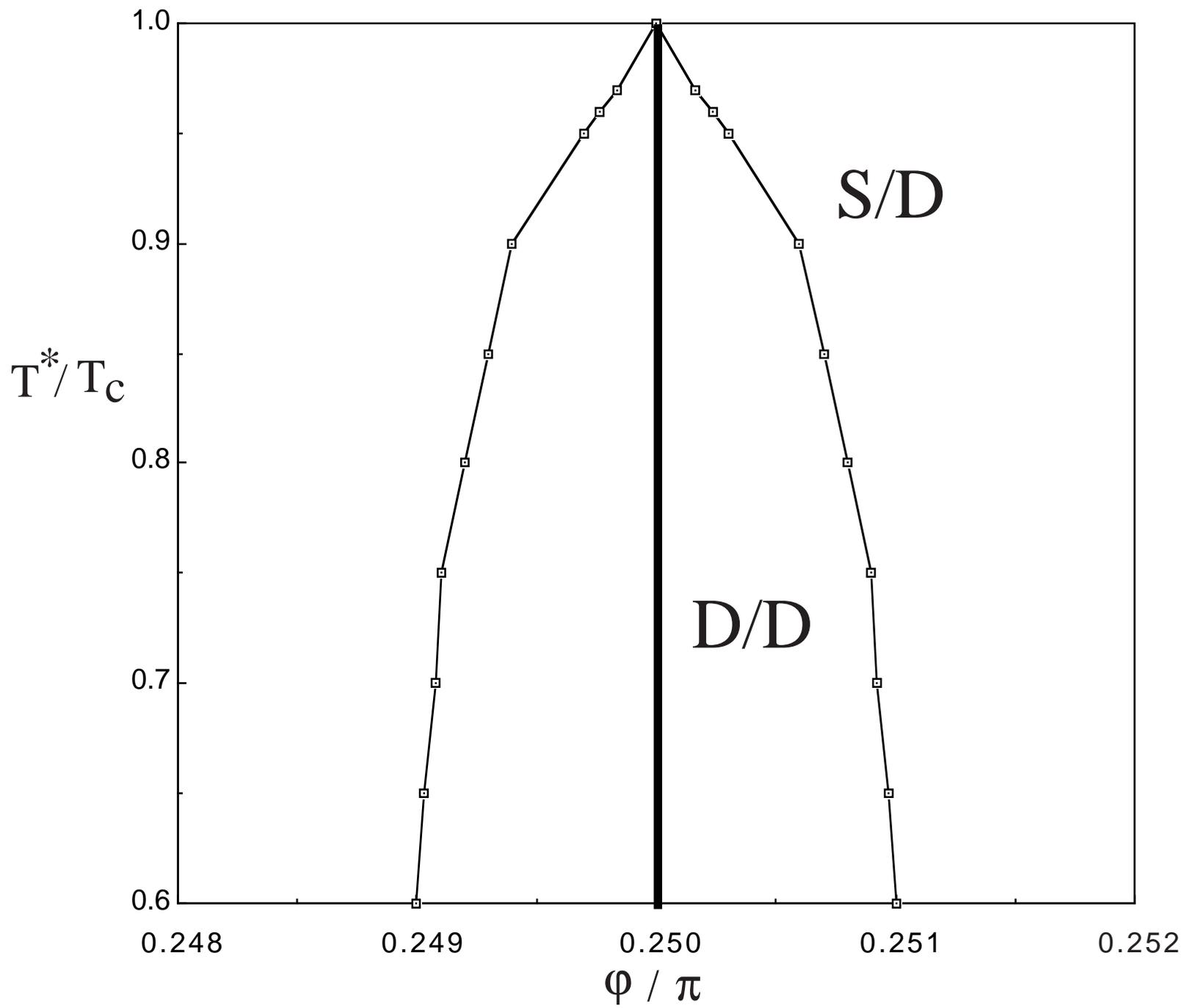
  {\bf Fig. 5.}  
  Phase diagram of surface states in the plane of $T$ and $\varphi$.
  Here $\delta = 0.15$, ${\tilde t}^2 = 8.0$ and ${\tilde r}^2 = 0.2$.
  The line is for the $D/D$ ($S/D$)-junction. (The ${\cal T}$-breaking 
  region for $D/D$-junction is invisible in this scale.)

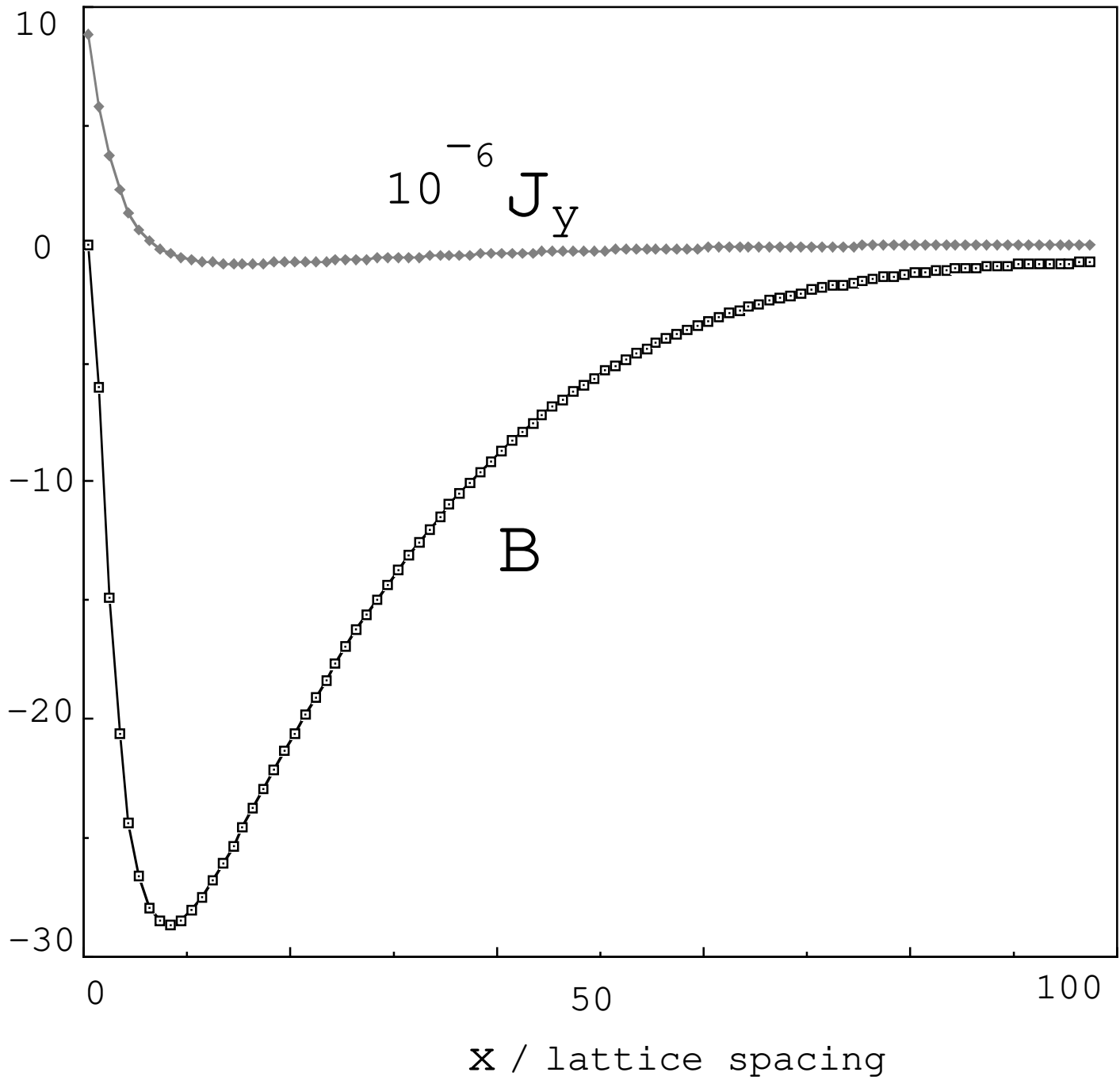
  {\bf Fig. 6.}  
  Distributions of the surface current $J_y$ (A/cm$^2$) 
  and the local magnetic field $B$ (G).
  Here $\delta = 0.15$, $T = 0.8T_c$, ${\tilde t}^2 = 8.0$,  
  ${\tilde r}^2 = 0.2$ and $\varphi = \pi/4$.
  
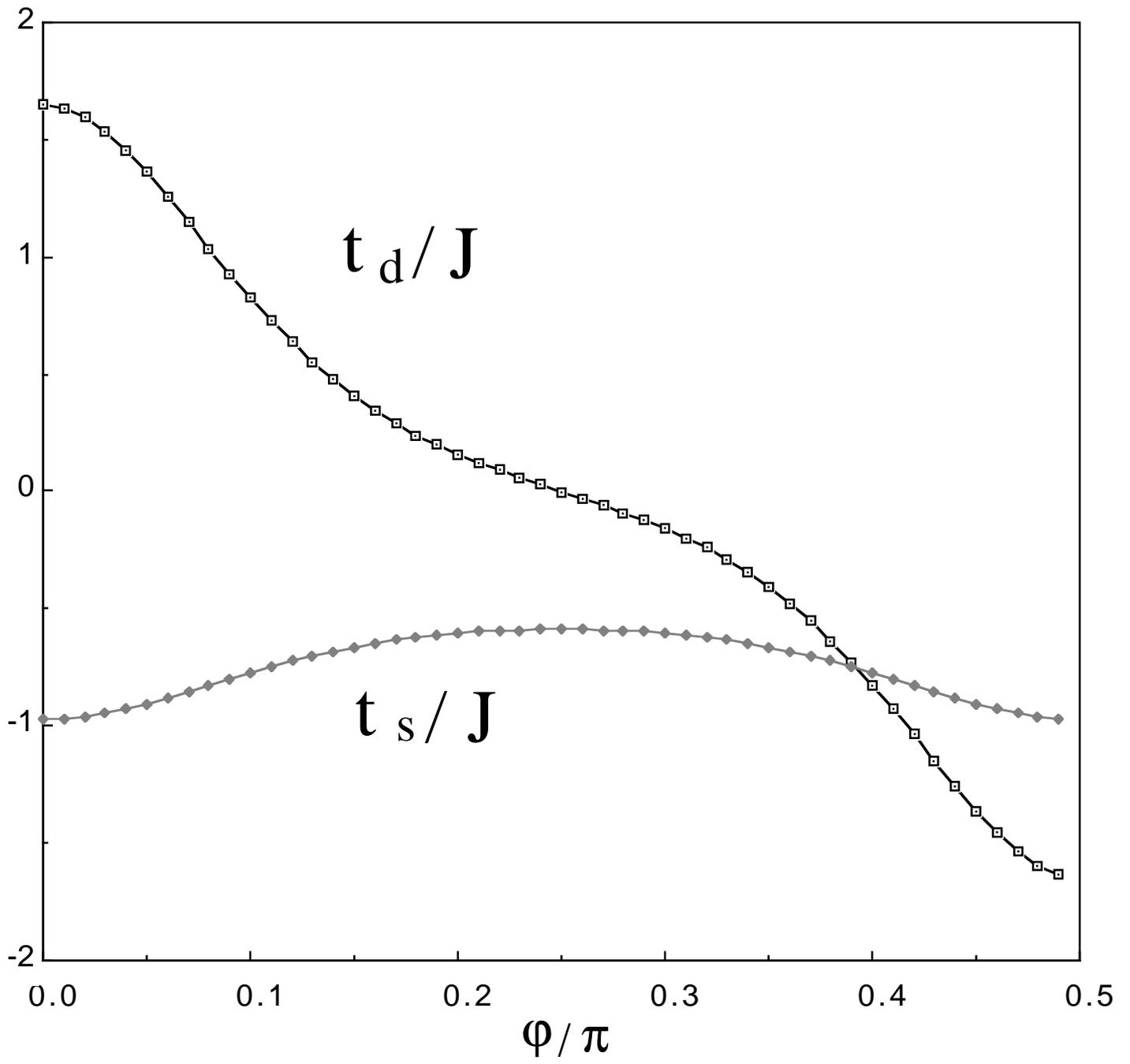
  {\bf Fig.7} 
  Numerical results of the coefficients in $F_I$ for $S/D$-junction.
  Here $\delta = 0.15$, $T = 0.8T_c$, ${\tilde t}^2 = 8.0$ 
  and ${\tilde r}^2 = 0.2$.
  
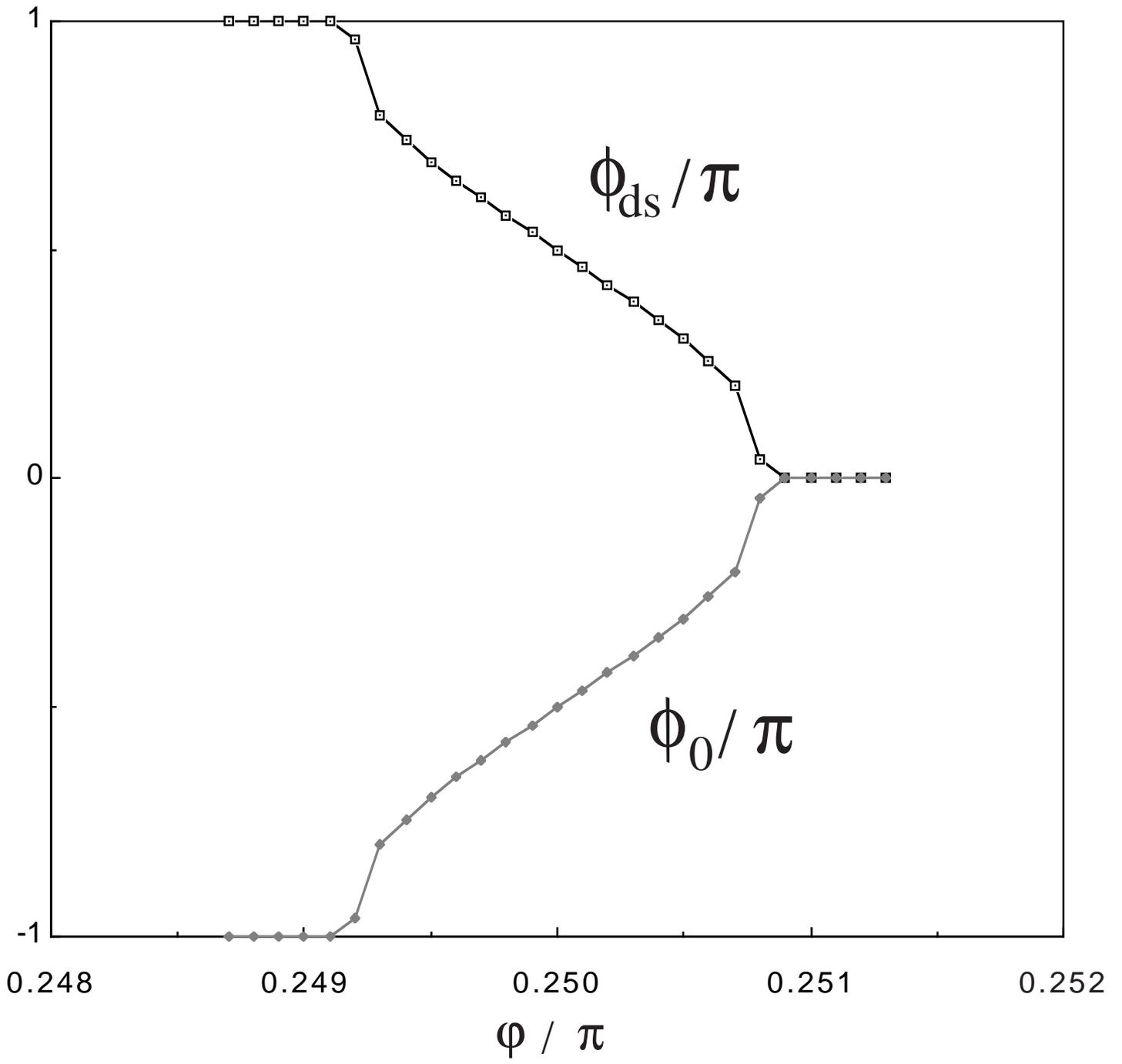
  {\bf Fig.8} 
  The relative phases between $\Delta_d$ and $\Delta_s$ ($\Delta_0$), 
  $\phi_{ds}$ ($\phi_0$) as a function of $\varphi$.


\begin{thebibliography}{99}
\bibitem{foot0}  In 2D Bose condensation in a usual sense would not occur. 
  However, there exists a crossover 
  temperature, $T_{BE}$, below which bosons are almost condensed to the 
  lowest energy state. The true condensation occurs with a small 
  three dimensionality which is always present in real systems.
\bibitem{foot1} In this state there is a (pseudo-) gap in the spin, but not 
  in the charge excitations.
\bibitem{foot2} In this theory the Fermi liquid state 
  is represented as an ordered state where holons are Bose condensed.  
  In this case the system has only fermionic 
  degrees of freedom (spinons), and thus represent a Fermi liquid state.
\bibitem{foot3} 
At very low temperature, of the order of $10^{-3}J$, $\alpha_s$ can 
  be negative. However, the GL theory could not be applied to such a 
  low $T$.
\end{thebibliography}
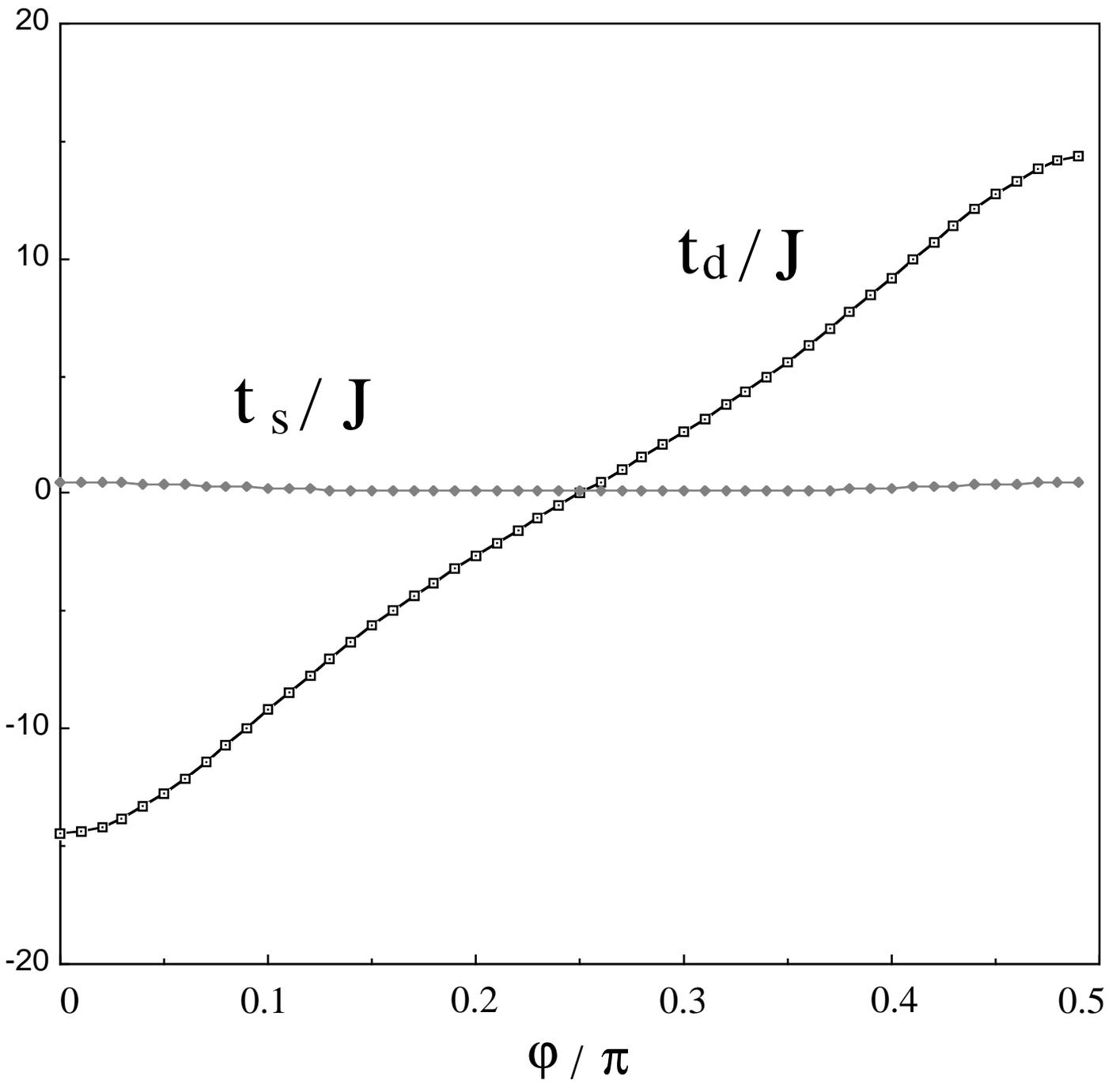
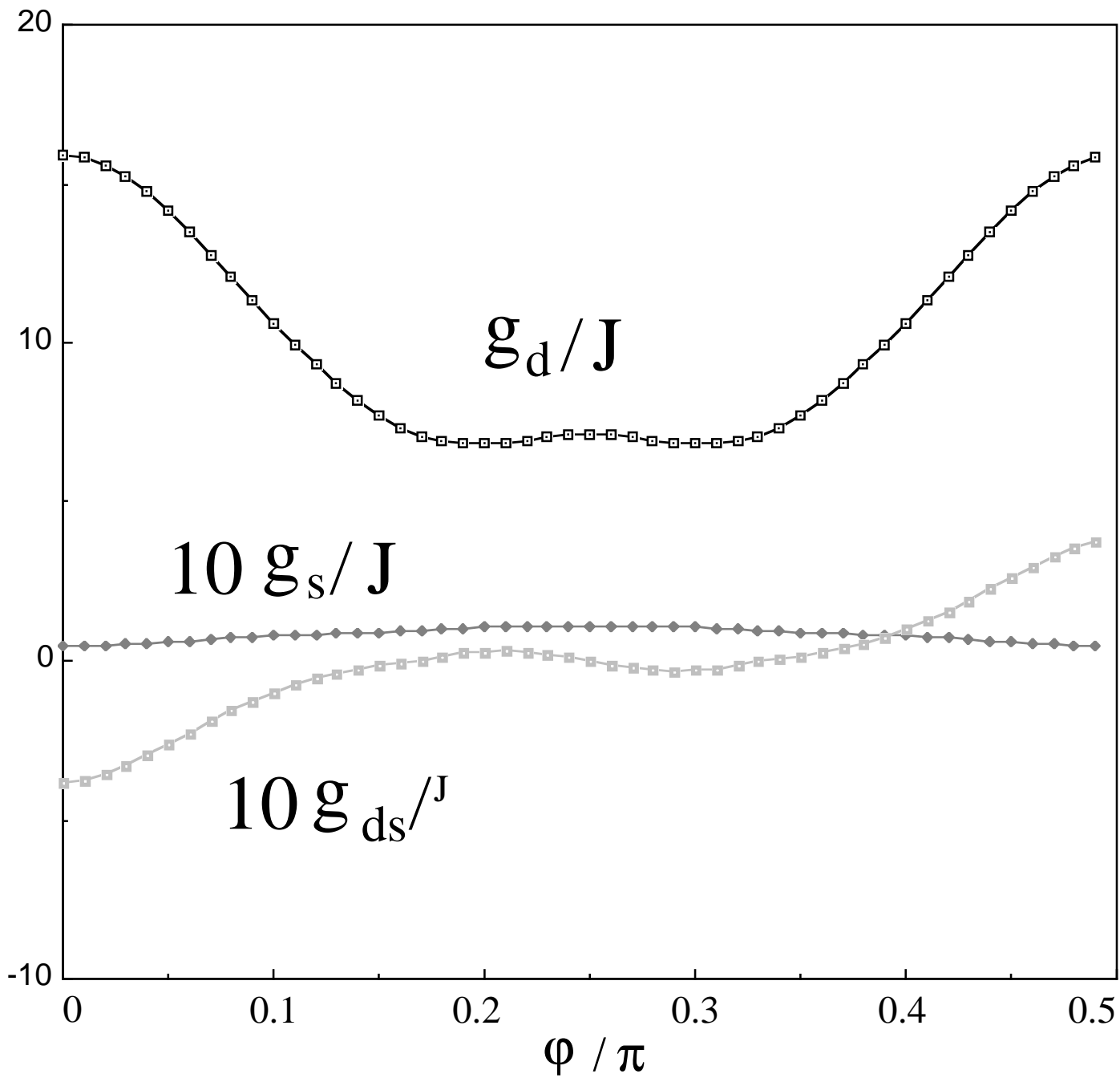
\end{document}